\documentclass{article}
\usepackage{spconf,amsmath,graphicx,amsfonts}

\title{A deep reinforcement learning approach to audio-based navigation in a multi-speaker environment}

\name{
    Petros Giannakopoulos$^{1}$,
    Aggelos Pikrakis$^{2}$,
    Yannis Cotronis$^{1}$}
\address{
    $^{1}$ National and Kapodistrian University of Athens, \{petrosgk,\;cotronis\}@di.uoa.gr \\
    $^{2}$University of Piraeus, pikrakis@unipi.gr}
\begin{document}

\copyright{2021 IEEE. Personal use of this material is permitted. Permission from IEEE must be obtained for all other uses, in any current or future media, including reprinting/republishing this material for advertising or promotional purposes, creating new collective works, for resale or redistribution to servers or lists, or reuse of any copyrighted component of this work in other works.}

\maketitle

\begin{abstract}
In this work we use deep reinforcement learning to create an autonomous agent that can navigate in a two-dimensional space using only raw auditory sensory information from the environment, a problem that has received very little attention in the reinforcement learning literature. Our experiments show that the agent can successfully identify a particular target speaker among a set of $N$ predefined speakers in a room and move itself towards that speaker, while avoiding collision with other speakers or going outside the room boundaries. The agent is shown to be robust to speaker pitch shifting and it can learn to navigate the environment, even when a limited number of training utterances are available for each speaker.
\end{abstract}
\section{Introduction}
\label{sec:intro}

We address the problem of environment navigation by an autonomous agent in a two-dimensional space, for the case where the sound sources are human speakers, using a Reinforcement Learning (RL) approach for training the agent. Our objective is to investigate whether an  agent can learn to navigate autonomously in a two-dimensional space, towards a target sound source (speaker) while simultaneously avoiding other sound sources (other speakers). An important constraint is that the agent is only allowed to use raw audio data in the form of a two-channel audio signal, in an attempt to simulate what a human listener would hear. For this purpose we created an environment in the Unity game engine \cite{juliani2018unity} which simulates a player, controlled by the agent, moving in a room where there also exist a number of speakers. The objective of the player is to move towards and eventually reach a specified target speaker, while at the same time avoiding coming into contact with the other speakers or going outside the room. The player receives stereo audio from the environment, processed by the Unity game engine to have pseudo-spatial two-dimensional information. Our experiments show that an agent trained with deep reinforcement learning, rewarded when reaching the target speaker and punished when bumping into other speakers or going outside the room, is able to solve the aforementioned episodic problem with a high degree of success.

\section{Related Work}
\label{sec:work}

So far, limited work has been published on audio-based virtual environment navigation for autonomous agents. An overview of the most related research efforts indicates that audio is usually treated as an additional  modality to image sensing. More specifically,   \cite{woubie2019autonomous} proposed an agent that uses audio information, in addition to visual information, to navigate a maze environment created in ViZDoom \cite{Kempka2016ViZDoom} and reach a goal. They found that the agent was able to reach the goal significantly faster and more reliably when using raw audio information from the environment in addition to visual information, compared to the case when  only visual information was processed by the agent. \cite{wang2014sound} developed a framework for sound localization and tracking of sound sources, along with visual information, for navigation of an autonomous agent in a virtual environment. Their system consists of a sound propagation model and sound localization model based on classical (non-DNN) algorithms. They show that an agent based on their system is successful in navigation, localization and collision avoidance in an environment with multiple sound sources. In \cite{lathuiliere2018deep}, the authors used Deep Reinforcement Learning for controlling the gaze of a robotic head based on audio and visual data from the virtual environment.

For the sake of completeness, we also provide a brief overview of related work in the neighboring disciplines of speaker localisation and speaker detection. This is because our agent is implicitly performing speaker detection while navigating and also because, even though it is not performing speaker localization, it can be considered as trying to move to the direction of the target speaker while avoiding obstacles, i.e., other speaker locations on the way. Speaker Identification \cite{furui200940} is a well studied field and in recent years, methods based on dynamic programming \cite{furtunua2008dynamic} and Machine Learning methods, particularly those based on Deep Neural Networks, have enjoyed success over classical methods \cite{sztaho2019deep}. Speaker localization from multiple audio sources in two and three-dimensional space is also a topic which has attracted a significant body of published work. Again, recent efforts focus on supervised Deep Learning methods for multiple speaker detection and localization problems (\cite{he2018deep}, \cite{wang2018robust}, \cite{chakrabarty2017multi}), as well as more general audio-source localization tasks (\cite{vera2018towards}, \cite{lathuiliere2018deep}) and performance improvement over  classic algorithms has been reported along with more flexible setups where fewer assumptions about the environment need to be made. 

Our approach is offering a simultaneous solution to the tasks of audio-based navigation and speaker detection and it is therefore different from the aforementioned efforts, i.e. audio is the only available modality for training an autonomous agent with deep reinforcement learning (and not an auxiliary one as in \cite{woubie2019autonomous, Kempka2016ViZDoom, wang2014sound,lathuiliere2018deep}) and, in addition, due to the need for controlling an avatar that navigates to and reaches a target speaker, algorithms that perform simultaneous speaker localization and detection are not readily applicable in our case.

\section{Agent and Environment Overview}
\label{sec:overview}

We now provide a description of our method and the virtual environment that we set up to apply and test the proposed approach. 

\subsection{Agent Architecture}
\label{ssec:architecture}

\begin{figure}[htbp]
\begin{minipage}[b]{1.0\linewidth}
  \centering
  \centerline{\includegraphics[width=7.0cm]{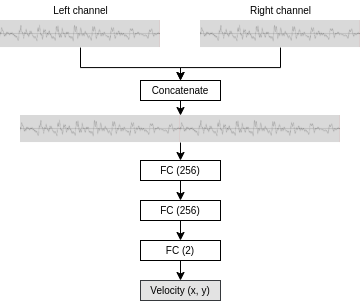}}
\end{minipage}
\caption{Agent's neural network architecture. The left and right channel of a stereo audio signal from the Unity environment are concatenated into a single vector. This vector is then fed into two fully connected hidden layers. The outputs of the network are the velocity of the agent-controlled player in the $x$ and $y$ axis, respectively.}
\label{fig:arch}
\end{figure}

Our agent is based on a combination of the Proximal Policy Optimization (PPO) reinforcement learning algorithm \cite{schulman2017proximal} with a small neural network consisting of two fully connected hidden layers. Our choice of reinforcement learning algorithm is not restrictive and other policy gradient optimization algorithms can be potentially suitable as well. Each hidden layer of the network has $256$ neurons and the output of the network is a fully connected layer consisting of $2$ neurons that estimate the velocity of the agent in the $x$ and $y$ directions, respectively. The input to the neural network is the the left and right channel s of the stereo audio signal produced by the Unity game engine, concatenated  into a single one-dimensional vector. In our experimental setup we retrieve the most recent $1024$ audio samples from the Unity engine's audio buffer for each channel. This means that, for a sampling frequency of $48$kHz, approximately, the most recent $21$ms  of the audio signal produced by the environment is the input frame to the agent at timestep $t$. The resulting vector, after the concatenation operation, has a length of $2048$ samples and it is fed as input to the network (Figure \ref{fig:arch}). We selected this particular architecture and input scheme after a coarse grid search over different fully-connected and 1D convolutional hidden-layer architectures, lengths of audio buffer and sampling frequencies.

\subsection{Environment}
\label{ssec:environment}

For our experimental study, we used the Unity editor to create a custom virtual environment that runs on the Unity game engine. We chose to create the environment in Unity after a search among other various known frameworks and platforms for creating virtual environments focused on training Reinforcement Learning Agents. The reason we eventually selected Unity was that it provided the easiest to use and most robust tools for creating environments containing audio sources with custom properties and also allowed for seamless capturing of the audio data from the simulated environment without the need for additional tools.

\begin{figure}[htbp]
\begin{minipage}[b]{1.0\linewidth}
  \centering
  \centerline{\includegraphics[width=8.5cm]{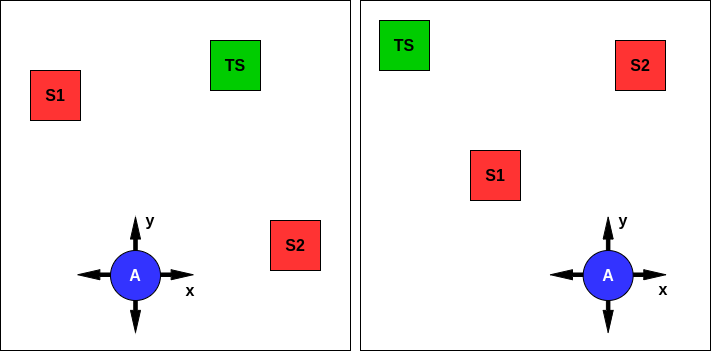}}
\end{minipage}
\caption{Top-down views of the environment used for the experiments. The views show two example state snapshots, one where agent $A$ has direct ``line-of-sight'' to $TS$ (left) and one where it does not (right). Agent $A$ can move along the $x$ and $y$ axes, while speakers $S1$, $S2$ and $TS$ are stationary. $TS$ is the target speaker and $S1$, $S2$ are other speakers in the room whose signals interfere with the signal produced by speaker $TS$. The agent $A$ must reach speaker $TS$ without bumping into (coming into contact) with speakers $S1$ and $S2$.}
\label{fig:playarea}
\end{figure}

Our test environment (Figure \ref{fig:playarea}) consists of  key components: 1) A rectangular room, 2)  speakers ($S1$, $S2$ and $TS$) inside the room, one of which is the target speaker (denoted $TS$) and 3) The agent ($A$) inside the room. Each speaker is a stationary Unity audio source which plays back an audio file containing an utterance for that speaker, randomly selected from a pool of utterances. When an utterance finishes, another one is randomly selected from the pool and played back. All audio sources share the same properties and are accordingly configured in the Unity editor. Specifically, the signal intensity (volume) of the audio source decreases linearly  with the distance from the source (linear volume roll-off). The maximum distance from which an audio source can still be heard is set so that all sources are audible across the entire room. The agent ($A$) can move inside the room along the $x$ and $y$ axes by adjusting its velocity vectors $v_x$ and $v_y$ over each axis. Each training episode begins with the agent appearing in a random location on the lower edge of the room and the speakers show up in random locations inside the room. The agent receives a positive reward of $+1.0$ when it reaches $TS$ and a negative reward of $-1.0$ when it crashes onto another speaker, i.e., $S1$ or $S2$ or when it moves outside the room boundaries. It also receives a small negative reward of $-0.001$ for every step it takes to discourage the agent from procrastinating. A training episode ends when the agent either reaches the target speaker or one of the other speakers or moves outside the room boundaries.

To facilitate reproducibility, we provide the Unity project files \footnote{https://github.com/petrosgk/AudioRL}. The Unity project includes all the required components for running the experiments: the Unity environment, a pre-trained agent and the dataset.

\subsection{Training the Agent}
\label{ssec:trainin}

The environment can be formalized as a Markov Decision Process (MDP) with the following characteristics: 1) The state $s$ at time $t$ (denoted $s_t$) consists of the two 1-D vectors, each representing the raw audio waveform from left and right channels of the stereo audio input from the environment, concatenated into a single 1-D vector, 2) Given $s_t$, the agent takes an action $a_t$, which consists of the normalized velocity vectors $v_x$, $v_y$, with $v_x \in [-1, 1]$ and $v_y \in [-1, 1]$, each applied to the respective axis of the agent-controlled player, 3) As a result of action $a_t$, the agent transitions to the next state $s_{t+1}$ and receives a reward $r \in [-1, 1]$.

We assume that there exists an optimal policy $\pi^*$, which, when followed by the agent, maximizes the cumulative reward $r$ achieved for a horizon of $T$ timesteps, where each timestep is an interaction with the environment. The agent's training objective is to find an as close as possible approximation $\pi_\theta$ to the optimal policy, such as $\pi_\theta \approx \pi^*$, where $\theta$ are the parameters of a neural network. To train the network parameters, the policy runs for $T$ timesteps and the collected samples are used for updating the policy gradient. In order to do this we first need to estimate the advantage at timestep $t$ of action $a_t$ in state $s_t$, as it was formulated in \cite{mnih2016asynchronous}:
\begin{equation}
    A_t(a_t, s_t) = \delta_t + \gamma\delta_{t+1} + \dots + \gamma^{T-t+1}\delta_{T-1},
\end{equation}
with
\begin{equation}
    \delta_t = r_t + \gamma V^{\pi}(s_{t+1}) - V^{\pi}(s_t),
\end{equation}
where
\begin{equation}
    V^{\pi}(s) = \mathbb{E}[\sum_{k=0}^\infty\gamma^kr_{t+k}]
\end{equation}
is the value estimate of state $s$ under policy $\pi$ and it is the mean expected reward, over $k$ timesteps, for following policy $\pi$ from state $s$.
In the above, $t \in [0,T]$ is the timestep index, $r_t$ is the reward at timestep $t$ and $\gamma \in (0, 1]$ is the future reward discount factor. The objective minimized by PPO is, according to \cite{schulman2017proximal}:
\begin{equation}
    L_t(\theta) = \mathbb{E}_t[L_t^{C}(\theta) - c_1 L_t^{VF}(\theta) + c_2 S[\pi_{\theta}](s_t)],
\end{equation}
where
\begin{equation}
    L^{C}(\theta) = \mathbb{E}_t[min(r_t(\theta)A_t, clip(r_t(\theta), 1-\epsilon, 1+\epsilon)A_t)]
\end{equation}
and
\begin{equation}
   L_t^{VF}(\theta) = (V_\theta(s_t) - V_t^{target})^2
\end{equation}
$c_1$, $c_2$ are coefficients, $S$ is an entropy bonus added to policy estimation, which incentivizes exploration, and $\epsilon$ is a hyperparameter. In our experiments we set $\gamma = 0.99$, $c_1 = 0.95$, $c_2 = 0.001$ and $\epsilon = 0.2$.

\section{EXPERIMENTS}
\label{sec:results}

\subsection{Testing Methodology}
\label{ssec:testing_methodology}

The dataset used for the experiments was created from three publicly available audiobooks, each read by one of the speakers (two male speakers and a female one). We first split each audiobook automatically in utterances, using a Voice Activity Detector (VAD) which labels the boundaries of each utterance. We then manually verify the VAD results and correct the boundaries of detected utterances where needed. We assume that the definition of an utterance is the linguistic one: \textit{"An uninterrupted chain of spoken or written language."} \cite{lexico_dictionaries}. This procedure resulted into a dataset of approximately $600$ utterances for each one of the three speakers.

At a next step, the pool of utterances of each speaker is split into a training partition and a testing partition. The overall training and testing sets are the union of these three training and testing partitions, respectively. The resulting training set consists of approximately $500$ utterances for each speaker, while the test set consists of around $100$ utterances per speaker. We then set one speaker as the target and train the agent for a total of $6$ million steps in the created  environment. Upon completion of the training stage, we evaluate the agent's performance by switching to the test set and letting the trained agent play $100$ episodes in the environment. For each episode we keep a record of success or failure. Success means that the agent reached the target speaker, while failure means that it reached another speaker or went outside the room boundaries. After the $100$ testing episodes are completed, we compute the agent's success rate and compare it with the corresponding success rate of a baseline agent performing a random action policy. This training-testing experiment is repeated three times by setting each time one of the three speakers as the Target Speaker. In the end, we average the success rates over all Target Speakers.

Furthermore, we evaluate the limits of generalization of our agent by training on only $1$ utterance per speaker and testing again on the full test set of $100$ utterances per speaker. In this way we can observe if our agent can still learn when very few training data are available for each speaker.

As a final test regarding the agent's generalization capabilities, we randomly shift the pitch frequency of each speaker by $4$ to $8$ percent for each test utterance and compute again the success rate. Note that the agent is only trained on the training set of $500$ utterances per speaker without pitch shifting (i.e., without data augmentation).

\subsection{Results}
\label{ssec:results}

\begin{figure}[htbp]
\begin{minipage}[b]{1.0\linewidth}
  \centering
  \centerline{\includegraphics[width=8.3cm]{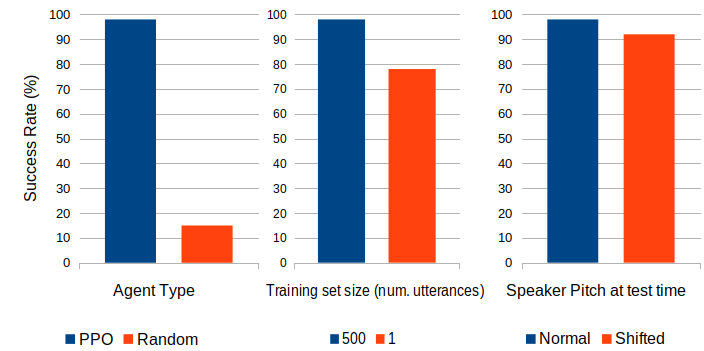}}
\end{minipage}
\caption{Left: Success rate of the PPO agent trained on environmental audio compared to a Random agent that explores the environment via a random walk. Middle: Success rate of the PPO agent after being trained on a set of $500$ utterances per speaker compared to the case where it is trained on only $1$ utterance per speaker. Right: Success rate of the PPO agent when the pitch of the speakers is randomly shifted between $4\%$ and $8\%$ during testing. The agent is trained on the original pitch of the speakers.}
\label{fig:results}
\end{figure}

Our findings are summarised in Figure \ref{fig:results}. Specifically:

a) The success rate of our reinforcement learning agent (based on PPO) on $100$ test episodes,  along with the success rate of an agent that moves randomly in the environment. Our agent is able to successfully reach the target speaker in $98$ out of $100$ played episodes ($98\%$ success rate). In contrast, the success rate of the random agent is $15\%$. This result can be interpreted as an indication that the proposed RL agent is capable of learning a representation of the environment that leads it to identify the target speaker and its position, based only on raw audio data. 

b) The success rate of the PPO agent when trained on the  training set of $500$ utterances per speaker and when it is only trained on $1$ utterance per speaker. Like before, the agent is again evaluated on the same test set of $100$ utterances per speaker. We saw that, when trained on the full training set, the agent has a $98\%$ success rate on the test set. This time, i.e., when trained on the minimal training set of just $1$ utterance per speaker it has a $78\%$ success rate on the test set. This result can be interpreted as an indication that the agent can generalize satisfactorily from very little training data.

c) The success rate of the PPO agent when the pitch of every speaker is randomly shifted between 4 and 8 percent during testing. It can be seen that, even though the agent was trained on the original pitch of the speakers, the success rate only exhibits a modest drop from $98\%$ to $93\%$, which leads us to believe the agent is also invariant, to a certain extent, to mild pitch shifting of the speakers.

\section{CONCLUSIONS}
\label{sec:conclusion}

In this work we investigated the performance of deep reinforcement learning in audio-based only navigation in a two-dimensional space containing  speakers as audio sources. An autonomous agent based on deep reinforcement learning was tasked with continuous control of a virtual entity moving in a room. The agent had to move the entity in the room in order to find and reach a particular speaker while not colliding with other speakers in the room. After the agent was trained in this task, its performance was evaluated using a test set of utterances not seen during training for each speaker. Our experiments showed that the agent was able to successfully learn this task from a state space consisting of time slices of each channel in the stereophonic raw environmental audio input. The agent was shown to generalize well to the unseen test set as it was able to complete the task with relatively high degree of success even when trained on a single utterance from each speaker. It was also shown to be robust to speaker pitch shifting not seen during training. As future work, we would like to examine scaling the experiments in larger and more complex environments with a larger number of different speakers and possibly other types of sound sources, e.g., environmental sounds.

\section{ACKNOWLEDGEMENTS}
\label{sec:acknowledgements}

The research project was supported by the Hellenic Foundation for Research and Innovation (H.F.R.I.) under the "1st Call for H.F.R.I. Research Projects to support Faculty Members \& Researchers and the Procurement of High-Cost Research Equipment Grant" (Project Number: 3449).

\bibliographystyle{main}
\bibliography{main}

\begin{thebibliography}{10}

\bibitem{juliani2018unity}
Arthur Juliani, Vincent-Pierre Berges, Esh Vckay, Yuan Gao, Hunter Henry,
  Marwan Mattar, and Danny Lange,
\newblock ``Unity: A general platform for intelligent agents,''
\newblock {\em arXiv preprint arXiv:1809.02627}, 2018.

\bibitem{woubie2019autonomous}
Abraham Woubie, Anssi Kanervisto, Janne Karttunen, and Ville Hautamaki,
\newblock ``Do autonomous agents benefit from hearing?,''
\newblock {\em arXiv preprint arXiv:1905.04192}, 2019.

\bibitem{Kempka2016ViZDoom}
Micha{\l} Kempka, Marek Wydmuch, Grzegorz Runc, Jakub Toczek, and Wojciech
  Ja\'skowski,
\newblock ``{ViZDoom}: A {D}oom-based {AI} research platform for visual
  reinforcement learning,''
\newblock in {\em IEEE Conference on Computational Intelligence and Games},
  Santorini, Greece, Sep 2016, pp. 341--348, IEEE,
\newblock The best paper award.

\bibitem{wang2014sound}
Yu~Wang, Mubbasir Kapadia, Pengfei Huang, Ladislav Kavan, and Norman~I Badler,
\newblock ``Sound localization and multi-modal steering for autonomous virtual
  agents,''
\newblock in {\em Proceedings of the 18th meeting of the ACM SIGGRAPH Symposium
  on Interactive 3D Graphics and Games}, 2014, pp. 23--30.

\bibitem{lathuiliere2018deep}
St{\'e}phane Lathuili{\`e}re, Benoit Mass{\'e}, Pablo Mesejo, and Radu Horaud,
\newblock ``Deep reinforcement learning for audio-visual gaze control,''
\newblock in {\em 2018 IEEE/RSJ International Conference on Intelligent Robots
  and Systems (IROS)}. IEEE, 2018, pp. 1555--1562.

\bibitem{furui200940}
Sadaoki Furui,
\newblock ``40 years of progress in automatic speaker recognition,''
\newblock in {\em International Conference on Biometrics}. Springer, 2009, pp.
  1050--1059.

\bibitem{furtunua2008dynamic}
Titus~Felix Furtun{\u{a}},
\newblock ``Dynamic programming algorithms in speech recognition,''
\newblock {\em Revista Informatica Economic{\u{a}} nr}, vol. 2, no. 46, pp. 94,
  2008.

\bibitem{sztaho2019deep}
D{\'a}vid Sztah{\'o}, Gy{\"o}rgy Szasz{\'a}k, and Andr{\'a}s Beke,
\newblock ``Deep learning methods in speaker recognition: a review,''
\newblock {\em arXiv preprint arXiv:1911.06615}, 2019.

\bibitem{he2018deep}
Weipeng He, Petr Motlicek, and Jean-Marc Odobez,
\newblock ``Deep neural networks for multiple speaker detection and
  localization,''
\newblock in {\em 2018 IEEE International Conference on Robotics and Automation
  (ICRA)}. IEEE, 2018, pp. 74--79.

\bibitem{wang2018robust}
Zhong-Qiu Wang, Xueliang Zhang, and DeLiang Wang,
\newblock ``Robust speaker localization guided by deep learning-based
  time-frequency masking,''
\newblock {\em IEEE/ACM Transactions on Audio, Speech, and Language
  Processing}, vol. 27, no. 1, pp. 178--188, 2018.

\bibitem{chakrabarty2017multi}
Soumitro Chakrabarty and Emanu{\"e}l~AP Habets,
\newblock ``Multi-speaker localization using convolutional neural network
  trained with noise,''
\newblock {\em arXiv preprint arXiv:1712.04276}, 2017.

\bibitem{vera2018towards}
Juan~Manuel Vera-Diaz, Daniel Pizarro, and Javier Macias-Guarasa,
\newblock ``Towards end-to-end acoustic localization using deep learning: From
  audio signals to source position coordinates,''
\newblock {\em Sensors}, vol. 18, no. 10, pp. 3418, 2018.

\bibitem{schulman2017proximal}
John Schulman, Filip Wolski, Prafulla Dhariwal, Alec Radford, and Oleg Klimov,
\newblock ``Proximal policy optimization algorithms,''
\newblock {\em arXiv preprint arXiv:1707.06347}, 2017.

\bibitem{mnih2016asynchronous}
Volodymyr Mnih, Adria~Puigdomenech Badia, Mehdi Mirza, Alex Graves, Timothy
  Lillicrap, Tim Harley, David Silver, and Koray Kavukcuoglu,
\newblock ``Asynchronous methods for deep reinforcement learning,''
\newblock in {\em International conference on machine learning}. PMLR, 2016,
  pp. 1928--1937.

\bibitem{lexico_dictionaries}
``Utterance: Definition of utterance by oxford dictionary,''
  https://www.lexico.com/en/definition/utterance,
\newblock Accessed: 22/10/2020.

\end{thebibliography}

\end{document}